\documentclass[aps,preprint,prd,showpacs,nofootinbib]{revtex4}

\usepackage{latexsym}
\usepackage{amsmath}
\usepackage{graphicx}
\usepackage{subfigure}
\usepackage{dcolumn}
\usepackage{bm}
\usepackage{amssymb}
\usepackage{latexsym}
\usepackage{ulem}
\usepackage{color}
\usepackage[colorlinks,linkcolor=magenta,anchorcolor=blue,citecolor=blue]{hyperref}

\def\be{\begin{equation}}
\def\ee{\end{equation}}
\def\ba{\begin{eqnarray}}
\def\ea{\end{eqnarray}}

\def\nn{\nonumber}
\def\lf{\left}
\def\rt{\right}

\begin{document}

\title{The slow expansion with nonminimal derivative coupling and its conformal dual  }

\author{Yong Cai$^{1}$\footnote{caiyong13@mails.ucas.ac.cn}}
\author{Yun-Song Piao$^{1}$\footnote{yspiao@ucas.ac.cn}}

\affiliation{$^1$ School of Physics, University of Chinese Academy of
Sciences, Beijing 100049, China}


\begin{abstract}

We show that the primordial gravitational wave with scale-invariant spectrum might emerge from a nearly Minkowski space, in
which the gravity is asymptotic-past free. We illustrate it with a
model, in which the derivative of background scalar field
nonminimally couples to gravity. We also show that  since
here the tensor perturbation is dominated by its growing mode,
mathematically our slowly expanding background is
conformally dual to the matter contraction, but there is no the
anisotropy problem.

\end{abstract}

\maketitle

\section{Introduction}\label{introduction}

The inflation paradigm is still the leading candidate of the
primordial universe, since it has elegantly solved several
problems of the hot big-bang cosmology
\cite{Guth:1980zm}\cite{Starobinsky:1980te}\cite{Linde:1981mu}\cite{Albrecht:1982wi}.
Maybe more attractively, inflation can generate primordial
perturbations,
which give a natural explanation for the origin of the large scale
structure and the CMB fluctuations. The nearly scale-invariant,
adiabatic, and Gaussian primordial curvature perturbation
predicted by slow-roll inflation is consistent with the recent
observations, such as Planck \cite{Ade:2015lrj}, see also
recent comments \cite{Linde:2014nna}\cite{Ijjas:2013vea}, while
the detection of primordial tensor perturbations
\cite{Starobinsky:1979ty}\cite{Rubakov:1982}, i.e., the primordial
gravitational waves (GWs), is still on the road. However,
inflation also suffers from the geodesic incompletion
problem \cite{Borde:2001nh}.

During inflation, the parameter $|\epsilon|=|{\dot H}|/H^2\ll 1$.
The evolution with $\epsilon\ll -1$ is called the slow
expansion, which may be asymptotically Minkowski in infinite past.
It was observed in \cite{Piao:2003ty} that such a spacetime might
be responsible for adiabatically producing the scale-invariant
curvature perturbation, which was implemented ghost-freely in
\cite{Liu:2011ns}\cite{Piao:2011bz}. This actually suggests a
scenario in which the scale-invariant adiabatic perturbation may
emerge from nearly flat Minkowski spacetime. The similar idea was
also proposed by Wetterich
\cite{Wetterich:2013aca}\cite{Wetterich:2014eaa} for different
motivation. The scale-invariant curvature perturbations can also
be obtained during the slow contraction ($\epsilon\gg1$), i.e.,
in ekpyrotic universe
\cite{Khoury:2001wf}\cite{Lehners:2011kr}, by applying adiabatic
ekpyrosis mechanism \cite{Khoury:2010gw}\cite{Joyce:2011ta},
though for ekpyrotic universe the entropic mechanism is actually
better to explain the observation
\cite{Li:2013hga}\cite{Fertig:2013kwa}\cite{Ijjas:2014fja}.

The detection of the primordial GWs is of great significance for
confirming general relativity (GR) and strengthen our confidence
in inflation. The scale-invariance of primordial GWs requires,
e.g.\cite{Piao:2011mq}
\ba a^2M_{P,eff}^2 & \sim & {1\over
(\tau_*-\tau)^2}, \label{Mp1}\\
 & or & (\tau_*-\tau)^4,
\label{Mp2}\ea
where $M_{P,eff}$ is the effective Planck scale and
$\tau=\int dt/a$. During inflation, $a\sim {1\over \tau_*-\tau}$,
so the spectrum is scale-invariant. While during the slowly
evolving, $a$ is approximately constant, hence the tensor
perturbation will be strongly blue, which is negligible on large
scale.

Nevertheless, GR might be modified when deal with the extremely
early universe, which will inevitably affect the primordial tensor
perturbations.
It was showed in \cite{Piao:2011bz} that if the effective Planck
scale grows rapidly during slow expansion, $M_{P,eff}^2\sim {1\over
(\tau_*-\tau)^2}$, which may be induced by the nonminimal coupling
of the scalar field to gravity, both curvature perturbation and
GWs produced may be scale-invariant.
Here, the slowly expanding
background is conformally dual to the inflation, see also
\cite{Qiu:2012ia}. It also further strengthens the argument
that the GWs amplitude does not necessarily determine the scale of
inflation \cite{Armendariz-Picon:2015dma}. Thus though this
result is actually a  reflection that the perturbations are
conformal invariant fully nonperturbatively\footnote{This could be understood as follows. The scale-invariance of GWs spectrum requires (\ref{Mp1}) or (\ref{Mp2}) to be satisfied. Since the perturbations are conformally invariant fully nonperturbatively, condition (\ref{Mp1}) or (\ref{Mp2}) generally indicates many different conformally dual backgrounds. Two special cases of them are the inflation (with $a^2\sim {1\over
(\tau_*-\tau)^2}$ and $M_{P,eff}^2=$constant) and the slow
expansion (with $M_{P,eff}^2\sim {1\over
(\tau_*-\tau)^2}$ and $a=$constant), both satisfy condition (\ref{Mp1}).},
e.g.\cite{Li:2015hga}\cite{Kubota:2011re},
it may offer us a different
angle of view to the inflation scenario itself and also the
primordial universe.

Recently, Ijjas and Steinhardt have proposed the anamorphic
universe \cite{Ijjas:2015zma}, in which the ekpyrosis is designed
as a conformally dual to the inflation, see also \cite{Li:2014qwa}
and the conflation \cite{Fertig:2015dva}. Moreover, it was
pointed out that the physics of these conformally dual backgrounds
are actually different when we see them from the matter point of
view.
The anamorphic
universe has no initial condition problem. This implies that
seeing inflation at its conformal angle of view might bring
fruitful perspective to its own issues, as well as
the physics of the primordial universe. Thus the relevant issues are
interesting for further study.

Recently, Wetterich has clarified how the scale-invariant
primordial perturbations arise in flat Minkowski space
\cite{Wetterich:2015ccd}, as in \cite{Piao:2011bz} base on the
scenario with (\ref{Mp1}).
Here, we will focus on that with (\ref{Mp2}),
which helps to better highlight the physics of primordial
perturbations and the role of conformal frames. We will see that
the scale-invariant primordial GWs may emerge from flat Minkowski
space, more interestingly, in which the gravity is asymptotic-past
free.

In Sec.II, we will give an overview of the slow expansion scenario.
After this, we will illustrate our thought with a model in
Secs.III and IV, in which the background field's derivative
nonminimally couples to gravity, which results in
$M_{P,eff}^2\sim (t_*-t)^4 \gg M_P^2$, so that the scale-invariant
primordial GWs may emerge from flat Minkowski space with
asymptotic-past free gravity.
In Sec.V, we find that though mathematically our
background is conformally dual to the matter contraction, there is
no the anisotropy problem, and  with the matter point of view, we
argue that our physical background is actually the expansion.

\section{Overview of slow expansion scenario}

\begin{figure}[htbp]
\includegraphics[scale=2,width=0.45\textwidth]{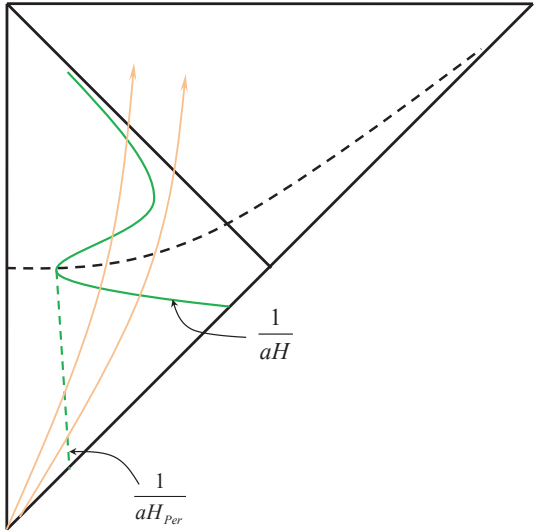}
\caption{The causal patch diagram of slow expansion scenario. The
universe is initially in Minkowski space, which then slowly
expands, reheats and evolves with big-bang cosmology. The black
dashed line is the reheating surface. The orange lines are
the comoving perturbation modes, which leave its comoving horizon
$1/(aH_{Per})$, but their wavelengths remain smaller than
$1/(aH)$. } \label{fig01}
\end{figure}

The slow expansion is the evolution with $\epsilon=-{\dot
H}/H^2\ll -1$. We may write it as \cite{Piao:2003ty} \be a\sim
(-t)^{-p}, \label{a1}\ee in which $0<p=1/|\epsilon|\ll 1$ is
constant, which is the scale solution, or
\cite{Liu:2011ns}\cite{Piao:2010bi} \be a\sim e^{{1\over
(-t)^n}}\simeq 1+{1\over (-t)^n}, \label{a2}\ee in which $n>0$ and
$\epsilon\sim -(-t)^n$, which is the Minkowski spacetime in
infinite past, and when ${1/(-t)^n}\sim 1$, the slow
expansion ends. It has been observed earlier in
\cite{Piao:2003ty} that such a spacetime might be responsible for
scale-invariant adiabatical perturbation, and after the end of the
slowly expanding phase, the universe may reheat and start to
evolve with standard cosmology. However, how to remove the ghost
instability still remains a challenging issue.  Recently, we have
implemented the corresponding scenarios ghost-freely in
\cite{Liu:2011ns} for (\ref{a2}) with $n=4$, and in
\cite{Piao:2011bz} for (\ref{a1}).

The physics of the origin of primordial perturbations is
illustrated as follows, see also the causal patch diagram in
Fig.\ref{fig01}. The perturbation mode with wavelength
$a\lambda\gg 1/H_{Per}$ will freeze and become the primordial
perturbation, otherwise it will oscillate inside $1/H_{Per}$. Here,
$1/H_{Per}$ is the sound horizon of the perturbations,  see Eq.(\ref{HT}) and (\ref{HR}) for its explicit expression. During
inflation, $1/(aH_{Per})$ nearly coincide with $1/(aH)$, which
suggests that the shape of perturbation spectrum is mainly set up
by the background evolution. However, during slow expansion,
$1/(aH_{Per})\ll 1/(aH)$, so the background is irrelevant with the
origin of primordial perturbation, as clarified recently also by
Wetterich in \cite{Wetterich:2015ccd}.

In (\ref{a2}), the physical meaning of $n$ is the ``slow-degree"
of expansion. The larger is $n$, the slower is the corresponding
expansion. It should be mentioned that $n=2$ is that of
the Galilean genesis
\cite{Creminelli:2010ba}\cite{Hinterbichler:2012yn}, so the
background (\ref{a2}) is also called the generalized genesis in
\cite{Nishi:2015pta}, in which it was clarified that the scale-invariant adiabatic perturbation can appear only for $n=4$.
When $n\ll 1$, (\ref{a2}) actually reduces to (\ref{a1}) with
$p=n$, since \be H={n\over (-t)^{1+n}}\simeq {p\over (-t)}\ee for
$n\ll 1$. When $n\gg 1$, the expansion is the slowest, (\ref{a2})
may be replaced with \be a\sim e^{1/(e^{-t})}\simeq 1+
1/(e^{-t}),\label{a3}\ee noting that initially $t\ll -1$, which
runs towards $t\simeq 0$. How the scale-invariant adiabatical
perturbation emerges from (\ref{a3}) was discussed in
\cite{Liu:2012ww}. The background (\ref{a3}) actually equals to
that in emergent scenario \cite{Ellis:2002we}, however, in which
it was implemented by introducing a positive curvature, so its
initial state is not flat Minkowski space, see also \cite{Bag:2014tta}.

As has been commented, the primordial GWs produced is generally
strong blue-tilt, which is negligible on large scale. However, if
$M_{P,eff}^2\sim {1\over (t_*-t)^2}$ is rapidly increasing during
slow expansion, which may be induced by the nonminimal coupling of
the scalar field to gravity, the primordial GWs may be
scale-invariant \cite{Piao:2011bz}.

\section{The model with nonminimal derivative coupling} \label{sec2}

\subsection{The Langrangian}

Here, we begin with \be S=\int dt d^3x \sqrt{-g}\left( {\cal L}_1+
{\cal L}_2\right)+{S}_{matter}, \label{L}\ee and \be {\cal L}_1=
-\,e^{4\phi/{\cal M}}X+{1\over {\cal M}^8}X^3-\alpha{\cal
M}^4\,e^{6\phi/{\cal M}}, \ee \be {\cal L}_2={M_P^2 \over
2}\left({\cal M}^8 /X^2+1\right)R+ {M_P^2 {\cal M}^8 \over
X^3}\left[-\left(\Box\phi\right)^2+\nabla_{\mu}\nabla_{\nu}\phi\nabla^{\mu}\nabla^{\nu}\phi\right],\label{L2}
\ee where $X=-{\nabla_\mu\phi\nabla^{\mu}\phi/ 2}$ and
$\Box\phi=g^{\mu\nu}\nabla_{\mu}\nabla_{\nu}\phi$, ${S}_{matter}$
is that of all components minimally coupling to the
metric, and ${\cal M}\ll M_P$, and $\alpha$ is constant.
Our (\ref{L}) is actually a subclass of Horndeski theory
\cite{Horndeski:1974wa}, which suggests that the equation of
motion is not higher than second order. The nonminimally
derivative coupling may be also
$G_{\mu\nu}\partial^{\mu}\phi\partial^\nu \phi$, which is also
interesting
\cite{deRham:2011by}\cite{Feng:2013pba}\cite{Sadjadi:2013psa}\cite{Goodarzi:2014fna}\cite{Heisenberg:2014kea}\cite{Feng:2014tka}\cite{Yang:2015pga}\cite{Zhu:2015lry}\cite{Qiu:2015aha}.

\subsection{The background of slow-expansion}\label{background}

Below we will derive the equation of motion for (\ref{L}), and
obtain the slowly expanding solution. The calculation is slightly
similar to that in \cite{Liu:2011ns}.

The Friedmann equation is
\ba 3M_P^2 H^2\left(1-\frac{15{\cal M}^8}{X^2}\right) & =  &
\rho_{eff} \nonumber\\ & = & -\,e^{4\phi/{\cal {\cal M}}}X+{5\over
{\cal {\cal M}}^8}X^3+\alpha {\cal M}^4\,e^{6\phi/{\cal M}} .
\label{rho0} \ea Here,
$\rho_{tot}=3M_P^2 H^2=\rho_{eff}+45M_P^2H^2{\cal M}^8/X^2>0$, and
$\rho_{eff}$ is that without the contribution from the derivative
coupling to gravity. We focus on the slowly expanding solution,
i.e., $H\simeq 0$ \cite{Piao:2003ty}, which requires
$\rho_{eff}\simeq 0$. In addition, we also require
$\rho_{eff,X}=0$ to avoid the divergence of the scalar spectrum, see Sec. \ref{scalar} for details. Both these two conditions fix the evolution of
$\phi$ as \be e^{\phi/{\cal M}}=\left({15\over 4}\right)^{1/4}
{1\over {\cal M}(t_*-t)}, \label{ephi}\ee \be {\dot \phi}={{\cal
M}\over (t_*-t)}, \label{dotphi}\ee and $\alpha={2\over
3\sqrt{15}}$, where initially $t$ is in negative infinity and runs
towards $t_*$. Thus only one adjustable parameter ${\cal M}$ in
(\ref{L}) is left, which, we will see, determines the amplitude of
primordial GWs.



\begin{figure}[htbp]
\includegraphics[scale=2,width=0.55\textwidth]{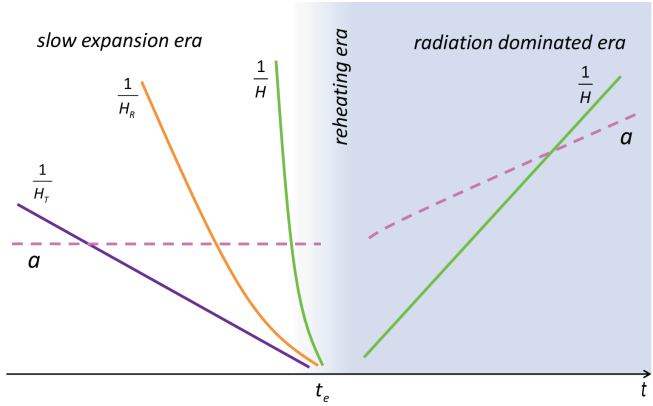}
\caption{The evolutions of the background, $1/H$, $1/H_{T}$ and
$1/H_{\cal R}$. There is a brief reheating era between the end of
the slow expansion era and the beginning of the radiation
dominated era. } \label{fig02}
\end{figure}

Using $\rho_{eff}\simeq0$, the evolution of $\dot H$ is given as
\ba 10{ {\cal M}^{8}M_P^2\over
X^2} {\dot H}  & \simeq & e^{4\phi/{\cal {\cal M}}}X-{1\over {\cal
{\cal M}}^8}X^3+ \alpha {\cal M}^4\,e^{6\phi/{\cal M}}+20{ {\cal
M}^{8}M_P^2\over X^3}H {\dot \phi}{\ddot \phi}\nonumber\\
& = & {24\over {\cal {\cal M}}^8}X^3+20{ {\cal M}^{8}M_P^2\over
X^3}H {\dot \phi}{\ddot \phi}. \label{dotH0} \ea After combining
Eq.(\ref{dotphi}), we have \be {\dot H}\sim {1\over {\cal M}^6
M_P^2} (t_*-t)^{-10}, \label{dotH}\ee which straightly gives \be
H\sim {1\over {\cal M}^6 M_P^2}(t_*-t)^{-9}. \ee The growing of
$H$ suggests the violation of the null energy condition. However,
the model is free of ghost instability, as will be showed in
Sec.\ref{sec3}.
The evolution of cosmological background is
\be a= a_0 e^{\int Hdt} \simeq a_0(1+{1\over {\cal M}^6
M_P^2}{1\over (t_*-t)^{8}})\simeq a_0.\label{a} \ee Thus initially
the universe is nearly Minkowski. Here, (\ref{a}) corresponds to
(\ref{a2}) with $n=8$. We plot the evolution sketch of the
background and $1/H$ in Fig.\ref{fig02}.


We see from (\ref{a}) that the condition of slow expansion is \be
{\cal M}^6 M_P^2 (t_*-t)^{8}\gg 1, \label{s-condition}\ee which
means $\epsilon\simeq -{\cal M}^6 M_P^2 (t_*-t)^{8}\ll -1$. When
the condition (\ref{s-condition}) is broken, the slowly expanding
phase ends, and ${\cal M}^6 M_P^2 (t_*-t)^{8}\simeq 1$ signals the
ending time $t_e$ \be t_*-t_{e}\simeq {1\over M_P^{1/4}{\cal
M}^{3/4} }. \label{te}\ee Hereafter, ${X/{\cal M}^4}\gg 1$, so GR
is recovered in (\ref{L}), the universe will evolve with the
standard cosmology.

\section{The power spectrum of primordial perturbations} \label{sec3}

\subsection{The primordial GWs}

We will calculate the primordial perturbations from nearly
Minkowski background (\ref{a}). Tensor mode $\gamma_{ij}$
satisfies $\gamma_{ii}=0$, and $\partial_i \gamma_{ij}=0$, and its
quadratic action for (\ref{L}) is \ba S^{(2)}=\int d\tau d^3x\,
{a^2 Q_T\over 8} \left[ \gamma_{ij}'^2-c_T^2(\vec{\nabla} \gamma_{ij})^2
\right],\label{action2} \ea where
\be Q_T =M_P^2\lf(1+{5{\cal M}^8\over X^2}\rt)\simeq 20  M_P^2
{\cal M}^4(t_*-t)^4, \label{QT} \ee and \be
c_T=\sqrt{\frac{1+{\cal M}^8/X^2}{1+5{\cal M}^8/X^2}} \simeq
\sqrt{1/5} \label{cT} \ee is the propagation speed of GWs. See
Appendix \ref{appendixa} for a notebook. Because $Q_T>0$ and $c_T^2>0$, there is no ghost instability.

In the momentum space, \be \gamma_{ij}(\tau,\mathbf{x})=\int
\frac{d^3k}{(2\pi)^{3} }e^{-i\mathbf{k}\cdot
\mathbf{x}}\sum_{\lambda=+,\times}\hat{\gamma}_{\lambda}(\tau,\mathbf{k})
\epsilon^{(\lambda)}_{ij}(\mathbf{k}), \label{hij}\ee where $
\hat{\gamma}_{\lambda}(\tau,\mathbf{k})=
\gamma_{\lambda}(\tau,k)a_{\lambda}(\mathbf{k})
+\gamma_{\lambda}^*(\tau,-k)a_{\lambda}^{\dag}(-\mathbf{k})$, the
polarization tensors $\epsilon_{ij}^{(\lambda)}(\mathbf{k})$
satisfy $k_{j}\epsilon_{ij}^{(\lambda)}(\mathbf{k})=0$ and
$\epsilon_{ii}^{(\lambda)}(\mathbf{k})=0$, and
$\epsilon_{ij}^{(\lambda)}(\mathbf{k})
\epsilon_{ij}^{*(\lambda^{\prime}) }(\mathbf{k})=\delta_{\lambda
\lambda^{\prime} }$, $\epsilon_{ij}^{*(\lambda)
}(\mathbf{k})=\epsilon_{ij}^{(\lambda) }(-\mathbf{k})$, the
commutation relation for the annihilation and creation operators
$a_{\lambda}(\mathbf{k})$ and
$a^{\dag}_{\lambda}(\mathbf{k}^{\prime})$ is $[
a_{\lambda}(\mathbf{k}),a_{\lambda^{\prime}}^{\dag}(\mathbf{k}^{\prime})
]=\delta_{\lambda\lambda^{\prime}}\delta^{(3)}(\mathbf{k}-\mathbf{k}^{\prime})$.

The equation of motion for $h_\lambda (\tau,k)$ is \be
u''+\left(c_T^2k^2-\frac{z_T''}{z_T} \right)u=0, \label{eom1} \ee
where $\gamma_{\lambda}(\tau,k)={u}(\tau,k)/{z_T}$ and $z_T= a {
\sqrt{Q_T} }/2$. Initially, the perturbations are deep inside
its own horizon $c_T/ H_T$, which means $c_T^2k^2 \gg
\frac{z_T''}{z_T}$. The GWs horizon $\sim 1/H_T$ is defined as
\be
H_T={1\over a}\sqrt{\frac{z_T''}{z_T}}\sim {1\over t_*-t}.\label{HT}
\ee
Here, since ${1/H\over c_T/H_T}\sim {\cal M}^6 M_P^2
(t_*-t)^{8}\gg 1$, even if the perturbations leave their own
horizon $c_T/H_T$, they remain inside $1/H$, see Fig.\ref{fig02}.
The primordial GWs spectrum is determined by $z_T=a\sqrt{Q_T}$.
Due to the rapid evolution of $Q_T$, the evolution
of background is now irrelevant to the origin of primordial
perturbation. The initial state of perturbation is the
Minkowski vacuum, \be u\sim \frac{1}{\sqrt{2c_T k} }e^{-i c_T
k\tau}. \label{initial1} \ee

When $c_T^2k^2 \ll \frac{z_T''}{z_T}$, i.e. the wavelength of
perturbation is far larger than its horizon $c_T/H_T$, the
solution of Eq.(\ref{eom1}) is given by \be u/z_T\sim C+\int
D{d\tau\over a^2 Q_T}, \label{uzT}\ee where $C$ is the constant
mode, while $D$ is the growing mode, $D\int {d\tau\over a^2Q_T}
\sim {1\over (t_*-t)^3}$, which will dominate the perturbation. We
have
\be |u|\simeq {1\over \sqrt{2c_T k}}\left(c_Tk
\tau_*-c_Tk\tau\right)^{-1}. \label{uT} \ee The power spectrum of
primordial GWs is \be {\cal P}_T^{1/2}={k^{3/2}\over
\sqrt{2\pi^2}} \sqrt{\sum_{\lambda=+,\times}
|\gamma_{\lambda}|^2}=\sqrt{4k^3\over \pi^2}{|u|\over a\sqrt{Q_T}}. \ee
Since the growing model dominates the perturbation, the amplitude
of perturbation will increase until the slow-expansion phase ends,
e.g.\cite{Liu:2011ns}. Therefore, the resulting spectrum of ${\cal P}_T$
should be calculated at $t_e$. Thus with Eq.(\ref{te}) and
(\ref{uT}), we have
\be {\cal P}_T^{1/2}\simeq \sqrt{1\over 10\pi^2}{1\over
c_T^{3/2}M_P {\cal M}^2 (t_*-t_e)^3} \simeq  \sqrt{5^{1/2}\over 2\pi^2}  \left({{\cal M}\over
M_P}\right)^{1/4}. \label{PT} \ee This indicates that the
primordial GWs is scale-invariant with the amplitude $({\cal M}/
M_P)^{1/4}$, and the only adjustable parameter $\cal M$ may be
fixed by the observation.

\subsection{The primordial scalar perturbations}\label{scalar}

The quadratic action for the curvature perturbation $\cal R$ is
\be S^{(2)}_{\cal R}= \int d\tau d^3x\, {a^4 Q_{\cal R}} \left[
{\dot{\cal R}}^2-{c_{\cal R}^2\over a^2}(\vec{\nabla} {\cal R})^2
\right],\label{action3} \ee where \be Q_{\cal R} \simeq 10{ {\cal
M}^8 M_P^2\over X^2}\sim M_P^2 {\cal M}^{4}(t_*-t)^{4}, \ee \be
c_{\cal R}^2\sim M_P^2 {\cal M}^{6}(t_*-t)^8, \ee see Appendix \ref{appendixa}
for a notebook. It should be mentioned that if $\rho_{eff,X}\neq
0$, we will have $Q_{\cal R}\sim (t_*-t)^{12}$ and $c_{\cal R}^2$
is constant, so that the spectrum of $\cal R$ will be strongly
red, which will make the amplitude of $\cal R$ diverge on the
largest scale.

Here, the sound speed changes rapidly. It is convenient to
redefine the conformal `time' as $dy=c_{\cal R}d\tau$ or
$d\tilde{y}=c_{\cal R}dt$, e.g.\cite{Khoury:2008wj}, which implies
\be \tilde{y}_*-\tilde{y} \sim {M_P {\cal M}^{3}\over
5}(t_*-t)^5.\label{y}\ee Thus the equation of motion for $\cal R$
is \be \frac{d^2v_{\cal R}}{dy^2}+\left(k^2-\frac{d^2z_{\cal
R}/dy^2}{z_{\cal R}} \right)v_{\cal R}=0, \label{eom2} \ee where
${\cal R}={v}_{\cal R}(\tau,k)/{z_{\cal R}}$ and $ z_{\cal R}= a {
\sqrt{c_{\cal R}Q_{\cal R}} }$.

Initially, the perturbations are deep inside its own horizon
$1/H_{\cal R}$, which means $k^2 \gg \frac{d^2z_{\cal
R}/dy^2}{z_{\cal R}}$. We have
\be {H}_{\cal R}={1\over
a}\sqrt{\frac{d^2z_{\cal R}/dy^2}{z_{\cal R}}}\sim {1\over
y_*-y}\sim {1\over (t_*-t)^5}.\label{HR}
\ee
Thus we see $ 1/H_T\ll 1/H_{\cal
R}\ll 1/H$, see Fig.\ref{fig02}. Since the evolution of ${H}_{T}$ is
distinguished from that of ${H}_{\cal R}$, the tilt of the $\cal
R$ spectrum must be different from that of GWs.

The initial state of perturbation is the Minkowski vacuum, \be
v_{\cal R}\sim \frac{1}{\sqrt{2 k} }e^{-i  ky}. \label{initial1}
\ee When $k^2 \ll \frac{d^2z_{\cal R}/dy^2}{z_{\cal R}}$, the
solution of Eq.(\ref{eom2}) is
\be |v_{\cal R}|\simeq {1\over \sqrt{2
k}}\left(ky_*-ky\right)^{1/5}, \label{u} \ee noting that $z_{\cal
R}\sim a M_P^{7/10} {\cal
M}^{11/10}(\tilde{y}_*-\tilde{y})^{4/5}$. Thus the power spectrum
of $\cal R$ is
\ba {\cal P}_{\cal R}^{1/2}& =& {k^{3/2}\over
\sqrt{2\pi^2}}\left|v_{\cal R}\over z_{\cal R}\right|\simeq
{1\over M_P^{5/2} {\cal
M}^{13/2}(t_*-t)^{9}}\left(ky_*-ky_e\right)^{6/5}\nonumber\\ &
\simeq & \left({{\cal M}\over M_P}\right)^{1/4}
\left(ky_*-ky_e\right)^{6/5}. \ea Thus the spectral index is
$n_{\cal R}-1=12/5$. Here, similar to GWs, the growing mode
dominates the perturbation, so the resulting spectrum of ${\cal
P}_T$ should be calculated at $t_e$.

\begin{figure}[htbp]
\includegraphics[scale=2,width=0.55\textwidth]{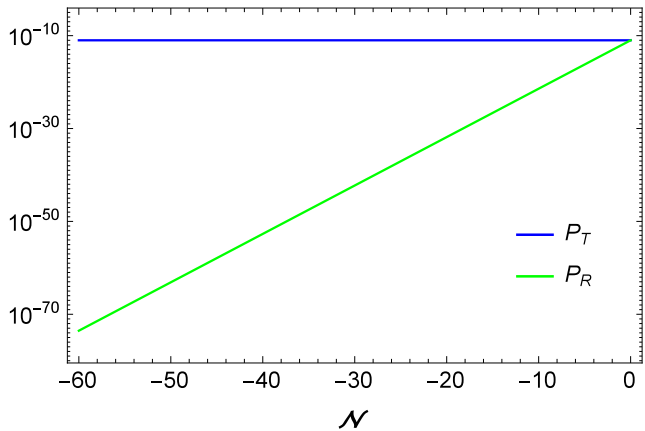}
\caption{${\cal P}_T$ and ${\cal P}_{\cal R}$ with respect to the
efolds number ${\cal N}=\ln(k/k_e)$. } \label{fig03}
\end{figure}

The amplitude of $\cal R$ spectrum should be same with that of
GWs, but since $n_{\cal R}-1=12/5$, the spectrum is blue-tilt, on
large scale the amplitude of $\cal R$ is negligible.
This can be shown as follows. We have $y_*-y_e\simeq
1/(aM_P^{1/4}{\cal M}^{3/4})$ in light of Eq.(\ref{y}), which just
equals to $1/k_e=1/(aH_e)$. The efolds number for the primordial
perturbations is defined as ${\cal N}=\ln(k/k_e)$. Thus ${\cal
P}_{\cal R}$ may be rewritten as \be {\cal P}_{\cal R} \simeq
\left({{\cal M}\over M_P}\right)^{1/2} e^{{12{\cal N}\over 5}},
\label{PR2} \ee where ${\cal N}<0$ since $k<k_e$. For ${\cal
N}>0$, the corresponding perturbation modes exits horizon after
$t_e$, thus will experience the evolution other than the slow
expansion. We assume ${\cal P}_T\sim 10^{-11}$, and plot ${\cal
P}_T$ and ${\cal P}_{\cal R}$ in Eqs.(\ref{PT}) and (\ref{PR2})
with respect to $\cal N$, respectively, in Fig.\ref{fig03}. We can see
that on far smaller scale, the amplitude of scalar perturbation is
same with that of GWs, but on large scales, ${\cal P}_{\cal R}$ is
negligible. Here, it is obvious that the adiabatic perturbation is
not able to be responsible for the CMB fluctuation and large scale
structure.

However, the curvature perturbation may also be induced by the
entropy perturbation from a light scalar field $\chi$  with
\be {\cal L}_{\chi}\sim -{1\over 2}e^{-{2\lambda\over \cal M}
\phi}(\partial \chi)^2,\label{nonchi}\ee  or nonminimally coupling
to the gravity ${\cal L}_{\chi}\sim -e^{-{2\lambda\over \cal M}
\phi}G_{\mu\nu}\partial^\mu \chi\partial^\nu \chi$
\cite{Feng:2013pba}\cite{Qiu:2013eoa}, where $\lambda$ is the
dimensionless constant. Defining $u_{\chi}=z_{\chi}\delta \chi$
and $z_{\chi}\sim a e^{-{\lambda\over \cal M}\phi}$, for
(\ref{nonchi}), we have the perturbation equation of $\delta\chi$
as \be
u_{\chi}''+\lf(k^2-\frac{z_{\chi}''}{z_{\chi}}\rt)u_{\chi}=0,\label{chi}
\ee where \be
\frac{z_{\chi}''}{z_{\chi}}\simeq\frac{\lambda^2-\lambda}{(
\tau_*-\tau)^2}. \ee
The power spectrum of $\delta\chi$ is $ {\cal
P}_{\delta\chi}={k^3\over2\pi^2}|\delta\chi|^2$.
Here, the mechanism
is similar to that applied to the ekpyrotic model, see Refs.
\cite{Li:2013hga}\cite{Fertig:2013kwa}\cite{Ijjas:2014fja}\cite{Li:2014yla}\cite{Levy:2015awa}
for the details, which may result in a local non-Gaussianity with
$f_{NL}\sim {\cal O}(1)$. When $\lambda=2$ or $-1$, ${\cal
P}_{\delta\chi}$ will be scale-invariant, see also \cite{Hinterbichler:2011qk}.
Then, we obtain \be u_{\chi}\simeq -{i\over \sqrt{2k}}\cdot{1\over
k(\tau_*-\tau)}\,, \ee for  $k(\tau_*-\tau)\ll 1$. Thus \be
|\delta\chi|={1\over \sqrt{2k^3}}\lf({15\over 4}\rt)^{
\lambda/4} \lf(M_p\over{\cal M}\rt)^{\lambda+1\over 4}{\cal M}\,.
\ee

After the slowly expanding phase ends, the universe
reheats, $\delta\chi$ may be convert to the curvature
perturbation. We follow Ref. \cite{Dvali:2003em}, and assume that
during reheating, the background field $\phi$ couples to ordinary
particles as $g^2\,\phi\,q\, q$, where $q$ represents the
ordinary particles, and $g$ is the coupling strength. Thus the
decay rate of $\phi$ is $\Gamma\sim g^2$. The coupling strength
can be $\chi$-dependent, \be g=g_0\lf(1+{\chi\over {\cal
M}_{\star}}\rt),\ee where ${\cal M}_{\star}$ is the scale with mass
dimension. After reheating, $\rho\sim T^4_{reh}$, and the
reheating temperature $T_{reh}\sim\sqrt{\Gamma}\sim g$. We have
\be {\cal R}={\delta\rho\over \rho}={4\delta\chi\over {\cal
M}_{\star} }\,,\ee noting ${\delta T_{reh}/ T_{reh}} ={\delta g/
g}={\delta\chi/{\cal M}_{\star} }$.

When $\lambda=-1$, we have \be
|\delta\chi|={1\over\sqrt{k^3}}\lf({1\over 15}\rt)^{1/4} {\cal
M}\,. \ee Then the power spectrum of $\cal R$ induced by
$\delta\chi$ is \be {\cal P}^{\delta\chi}_{\cal
R}={k^3\over2\pi^2}|{\cal R}|^2={8\over\sqrt{15}\pi^2}\lf({{\cal
M}\over{\cal M}_{\star} }\rt)^2\,. \ee Thus with (\ref{PT}), we have \be
r={{\cal P}_{T}\over {\cal P}^{\delta\chi}_{\cal
R}}={5\sqrt{3}{\cal M}_{\star}^2\over16 \sqrt{{\cal M}^3M_P}} \,. \ee
Recent observations suggested $r<0.1$ \cite{Ade:2015lrj}, which puts the bound
${\cal M}_{\star}< 0.43 M_P^{1/4}{\cal M}^{3/4}$. Moreover, by
requiring $\lambda$ to slightly deviate from $-1$, we will obtain
a nearly scale-invariant scalar spectrum with slight red tilt.

\subsection{The Minkowski space with asymptotic-past free gravity }

As was showed in Sec.\ref{background}, the initial universe is in a flat
Minkowski space. The initial background is not spoiled by
the perturbations, since the average square of the amplitude of
$\cal R$ in infinite past is \be <{\cal R}^2>={1\over
8\pi^3}\int^{aH}_{aH/e}|{\cal R}|^2d^3k\simeq {H^{12/5}\over {\cal
M}^{11/10}M_P^{13/10}}\longrightarrow 0.  \ee

\begin{figure}[htbp]
\includegraphics[scale=2,width=0.85\textwidth]{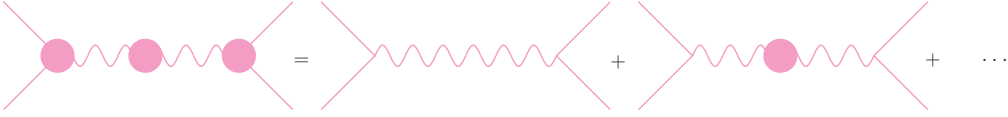}
\caption{ One particle reducible graphs for the gravity
interaction, in which the solid circles denote the full set of
radiation corrections to the vertex function and the graviton
propagator.  } \label{fig04}
\end{figure}

In initial Minkowski space with $a=a_0$, see (\ref{a}), the cubic
action of tensor perturbation \cite{Gao:2011vs} and, e.g., the
interaction between it and the Dirac field \cite{Feng:2015dxa},
are
\begin{eqnarray}\label{S3}
S^{(3)}=\int d^4x {a_0 c_T^2 Q_T\over
4}\left(\gamma_{ik}\gamma_{jl}-{\gamma_{ij}\gamma_{kl}\over
2}\right)\partial_k\partial_l \gamma_{ij},
\end{eqnarray}
\begin{eqnarray}\label{eq:int-action-massless}
{S}_{\psi{\bar \psi}\gamma}^{(3)} & = & \int d^4x {a^2_0\over 2}T^{ij}
\gamma_{ij}\nonumber\\
& = & i\int d^4x~\frac{
a_0^2}{8}\left(\bar{\psi}\gamma^i(\partial^j\psi)+\bar{\psi}\gamma^j(\partial^i\psi)
 -(\partial^i\bar{\psi})\gamma^j\psi-(\partial^j\bar{\psi})\gamma^i\psi\right)\gamma_{ij},
\end{eqnarray}
respectively. We redefine $\gamma_{ij}$ as
$M_{P,eff}\gamma_{ij}/2$ \cite{Donoghue:1994dn}, and write
(\ref{S3}) and (\ref{eq:int-action-massless}) as
\begin{eqnarray}\label{S3r} S^{(3)}=\int a_0 d^4x { 2\over
M_{P,eff}}\left(\gamma_{ik}\gamma_{jl}-{\gamma_{ij}\gamma_{kl}\over
2}\right)\partial_k\partial_l \gamma_{ij},
\end{eqnarray}
\begin{eqnarray}\label{eq:int-action-masslessr}
{S}_{\psi{\bar \psi}\gamma}^{(3)} & = & \int a_0^2 d^4x {1\over
M_{P,eff}}T^{ij} \gamma_{ij},
\end{eqnarray} where $M_{P,eff}=c_T\sqrt{Q_T}$.
The strength of the gravitational interaction is determined
by a set of one particle reducible graphs, see Fig.\ref{fig04}.
Thus, after neglecting the tensor index, we may write the renormalizated effective Newton
constant $G_{N,ren}$ as  \be
i{G_{N,ren}\over p^2}={i\over M_{P,eff}^2p^2}+{i\over
M_{P,eff}^2p^2}(loop) {i\over M_{P,eff}^2p^2} +\cdots. \ee
Initially, $M_{P,eff}=\sqrt{c_T^2Q_T} \sim (t_*-t)^2$ is infinite
large, which implies $G_{N,ren}=0$.
Thus from $t<t_e$, if we back to the infinite past, the
gravitational force will fade gradually and disappear eventually.
This suggests that the gravity is asymptotic-past free.

However, after $t\simeq t_e$, \be {X\over {\cal M}^4}\simeq
{1\over {\cal M}^2(t_*-t_e)^2}=\left({M_P\over {\cal
M}}\right)^{1/2}\gg 1 \ee we will have $M_{P,eff} \simeq M_P$, GR
is recovered, hereafter the universe will evolve with the standard
cosmology. Thus the asymptotic-past freedom of gravity is not
conflicted with our current observations.



\section{See it in Einstein frame}\label{appb}

\subsection{The Langrangian}

In principle, for the action with  nonminimal coupling to
gravity, it is always possible to rewrite it to the
Einstein-Hilbert's, in which the Ricci scalar is minimally
coupled.

We rescale the metric as \be g_{\mu\nu}={\cal A}^2({\hat X})
\hat{g}_{\mu\nu},\label{conformal} \ee
which implies \be g=\hat{g} {\cal A}^8,\qquad
R=\frac{\hat{R}}{{\cal A}^2}-\frac{6{\hat \nabla}_{\mu}{\hat
\nabla}^{\mu}{\cal A} }{{\cal A}^3}, \ee \be
\Box\phi=\frac{\hat{\Box}\phi}{{\cal A}^2}+\frac{2{\hat
\nabla}_{\mu}{\cal A}{\hat \nabla}^{\mu}\phi }{{\cal A}^3},\qquad
X=\frac{\hat{X}}{{\cal A}^2}. \ee Thus for (\ref{L}), we have \be
S=\frac{M_p^2}{2}\int d^4x\sqrt{-{\hat g}}{\cal A}^2(1+\frac{{\cal
M}^8}{{\hat X}^2}{\cal A}^4) {\hat R} +\cdots, \label{L11}\ee
which is Einstein-Hilbert style requires ${\cal
A}^2(1+\frac{{\cal M}^8}{{\hat X}^2}{\cal A}^4)=1$. This gives
\be{\cal A}=\frac{{\hat X}^{1/3} }{ {\cal M}^{4/3} }={X\over {\cal
M}^4}\label{A2}, \ee since $\frac{{\cal M}^8}{{\hat X}^2}{\cal
A}^4\gg 1$ during the slow expansion.
The line element is $d\hat{s}^2={\cal
A}^{-2}ds^2=\hat{g}_{\mu\nu}dx^{\mu}dx^{\nu}$. It's convenient to
redefine $d{\tilde t}={\cal A}^{-1}dt$ and $\tilde{a}={\cal
A}^{-1} a$, which make us back to the Einstein frame
$d\tilde{s}^2=d\hat{s}^2=\tilde{g}_{\mu\nu}d\tilde{x}^{\mu}d\tilde{x}^{\nu}=-d\tilde{t}^2+\tilde{a}^2d\mathbf{x}^2$,
in which (\ref{L11}) becomes \ba S_E&=&\frac{M_P^2}{2}\int
d\tilde{t} d^3\mathbf{x} \sqrt{-\tilde g} \Big({\tilde R}
-2\alpha \frac{{\tilde X}^{4/3} }{M_P^2{\cal M}^{4/3}}e^{6\phi/{\cal M}}-2\frac{{\tilde X}^{5/3} }{M_P^2{\cal M}^{8/3}}e^{4\phi/{\cal M}}\nn\\
&\,&+2\frac{{\tilde X}^{7/3} }{M_P^2{\cal
M}^{16/3}}-2\frac{({\tilde\nabla}_{\mu}{\tilde\nabla}^{\mu}\phi)^2}{\tilde
X} -\frac{4}{9}\frac{({\tilde\nabla}_{\mu}{\tilde
X}{\tilde\nabla}^{\mu}\phi)^2}{{\tilde X}^3}
-2\frac{{\tilde\nabla}_{\mu}{\tilde\nabla}^{\mu}{\tilde X}}{{\tilde X}}\nn\\
&\,&-2\frac{{\tilde\nabla}_{\mu}{\tilde\nabla}^{\mu}{\tilde
X}}{{\cal M}^{8/3}{\tilde X}^{1/3} }
+\frac{4}{9}\frac{{\tilde\nabla}_{\mu}{\tilde
X}{\tilde\nabla}^{\mu}{\tilde X}}{{\tilde X}^2}
+\frac{4}{3}\frac{{\tilde\nabla}_{\mu}{\tilde
X}{\tilde\nabla}^{\mu}{\tilde X}}{{\cal M}^{8/3}{\tilde X}^{4/3}
}-\frac{4}{3}\frac{{\tilde\nabla}_{\mu}\phi{\tilde\nabla}^{\mu}{\tilde
X}{\tilde\nabla}_{\nu}{\tilde\nabla}^{\nu}\phi}{{\tilde X}^2}
\nn\\
&\,& -\frac{8}{3}\frac{{\tilde\nabla}^{\mu}{\tilde
X}{\tilde\nabla}^{\nu}\phi{\tilde\nabla}_{\nu}{\tilde\nabla}_{\mu}\phi}{{\tilde
X}^2}
+2\frac{{\tilde\nabla}_{\mu}{\tilde\nabla}_{\nu}\phi{\tilde\nabla}^{\mu}{\tilde\nabla}^{\nu}\phi}{\tilde
X} \Big). \label{ml}\ea

\subsection{The background}

The evolutions of ${\tilde H}={d{\tilde a}\over d\tilde t}/{\tilde
a}$ and $\dot{\tilde H}$ are
\ba 3{\tilde H}^2M_P^2&=&-\alpha e^{6\phi/{\cal M}}\frac{{\tilde X}^{4/3}}{15{\cal M}^{4/3} }+e^{4\phi/{\cal M}}\frac{{\tilde X}^{5/3}}{15{\cal M}^{8/3} }-\frac{ {\tilde X}^{7/3}}{3{\cal M}^{16/3} }\nn\\
&\,&-2M_P^2 \frac{{\tilde H}\dot{\phi}\ddot{\phi} }{\tilde
X}-4M_P^2 \frac{{\tilde H}\dot{\phi}\ddot{\phi} }{5{\cal
M}^{8/3}{\tilde X}^{1/3} }
-2M_P^2 \frac{\ddot{\phi}^2 }{3\tilde X}\nn\\
&\,&-2M_P^2 \frac{\ddot{\phi}^2 }{45{\cal M}^{8/3}{\tilde X}^{1/3}
}-4M_P^2 \frac{\dot{\phi}\dddot{\phi} }{15{\cal M}^{8/3}{\tilde
X}^{1/3} }\,\,,\label{H2MP2} \ea \ba \dot{\tilde H}&=&e^{6\phi/{\cal
M}}\frac{2\alpha {\tilde X}^{4/3} }{15{\cal
M}^{4/3}M_P^2}+e^{4\phi/{\cal M}}\frac{ {\tilde X}^{5/3} }{15{\cal
M}^{8/3}M_P^2}
+\frac{ {\tilde X}^{7/3} }{15{\cal M}^{16/3}M_P^2}\nn\\
&\,&+\frac{{\tilde H}\dot{\phi}\ddot{\phi} }{\tilde
X}+\frac{2{\tilde H}\dot{\phi}\ddot{\phi} }{5{\cal M}^{8/3}{\tilde
X}^{1/3} }+\frac{\ddot{\phi}^2}{\tilde
X}+\frac{4\ddot{\phi}^2}{45{\cal M}^{8/3}{\tilde
X}^{1/3}}-\frac{\dot{\phi}\dddot{\phi} }{3\tilde
X}+\frac{2\dot{\phi}\dddot{\phi} }{15{\cal M}^{8/3}{\tilde
X}^{1/3}}\,\,, \ea respectively.
Note that only in this subsection a dot denotes $d/d\tilde{t}$.
Here, both equations involve the higher-order derivatives of
$\phi$. Thus straightly acquiring the solution of background is
difficult. However, it is convenient to calculate it by using the
conformal relation (\ref{A2}). We have \be {\cal A}={X\over {\cal
M}^4}={1\over { 2} {\cal M}^2(t_*-t)^{2}}, \ee where
Eq.(\ref{dotphi}) is applied. Then noticing $d{\tilde t}={\cal
A}^{-1}dt$, we have ${\tilde t}_*-{\tilde t}=2{\cal
M}^2(t_*-t)^3/3$. Thus the background is \be \tilde{a}(\tilde
t)={\cal A}^{-1}a_0(1+{1\over {\cal M}^6 M_P^2}{1\over
(t_*-t)^{8}})\simeq a_0{\cal M}^{2/3}({\tilde t_*}-{\tilde
t})^{2/3}, \ee  which is the matter contraction
\cite{Wands:1998yp}\cite{Finelli:2001sr}, also
\cite{Starobinsky:1979ty}.




\subsection{The physical background}

The frame that the matter minimally couples to the metric may be
called the matter frame. The conformally dual models can be
distinguished when we see from  the matter point of view, as has been
clarified in \cite{Domenech:2015qoa}, or by applying the
Weyl-invariants \cite{Ijjas:2015zma} \be \Theta_m=\lf(H+\frac{\dot
m}{m} \rt)M_{P,eff}^{-1}, \ee \be
\Theta_P=\lf(H+\frac{\dot{M}_{P,eff}}{M_{P,eff}}
\rt)M_{P,eff}^{-1} \ee where $\Theta_m$ defines a physical ruler
measuring the evolution of background, the physical ruler is
comprised of particles with mass $m$, $S_{matter}=\int m ds$, and
its length scale is set by the Compton wavelength
$\lambda_{Compton}\sim 1/m$ of the particle, $\Theta_m>0$ signals
that the background felt by the matter is expanding,  otherwise it is
contracting, while $\Theta_P=({{\dot z_T}/ z_T})M_{P,eff}^{-1}$
measures the evolutive behavior of $1/H_T$, and so the primordial
GWs, the conformal invariance of perturbations suggests that
conformally dual models have same $\Theta_P$. The inflation
corresponds to $\Theta_m>0$, $\Theta_P>0$, while the ekpyrosis is
$\Theta_m<0$, $\Theta_P<0$. Recently, the anamorphic universe
proposed in \cite{Ijjas:2015zma} corresponds to $\Theta_m<0$,
$\Theta_P>0$.

We focus on the slow expansion model proposed here. We have
$\dot{m}=0$, since the matter minimally couples to the metric, and
the effective Planck mass \be M_{P,eff}=\sqrt{c_T^2Q_T}
=M_P\sqrt{1+\frac{{\cal M}^8}{X^2}} \sim (t_*-t)^2. \label{Mpeff}
\ee Thus $\Theta_m\sim H>0$ and $\Theta_P\sim
\frac{\dot{M}_{P,eff}}{M_{P,eff}}<0$, see also
\cite{Fertig:2015dva} for conflation. $\Theta_m>0$ suggests our
cosmological background is actually a physical expansion when
measured relative to a physical ruler ${\tilde\lambda}_{Compton}$,
regardless of the frames. Actually, it's easy to check that
in Einstein frame,
${\tilde\lambda}_{Compton}\sim1/\tilde
m\sim(\tilde{t}_*-\tilde{t})^{2/3}$ is actually contracted
slightly faster than $\tilde a$, in which $S_{matter}=\int
\tilde{m}d{\tilde s}$ and $\tilde{m}=m\cal A$. Thus the universe
is a physical expansion when compared with
${\tilde\lambda}_{Compton}$.


The matter contraction scenario is still censured for its anisotropy
problem. In our conformally dual model, the contribution of the
anisotropy is \be \sigma^2\sim (t_*-t)^{-8}, \ee see Appendix \ref{appendixb}
for the details, while $\rho_{tot}\simeq
3M_P^2H^2\sim(t_*-t)^{-18}$ grows far faster than $\sigma^2$.
Therefore, our model does not suffer from the anisotropy problem
provided $\sigma^2\ll \rho$ initially. However, in the Einstein frame, the evolution of the anisotropy is
\be {\tilde\sigma}^2 \sim
1/{\tilde a}^{6}\sim {1\over ({\tilde t}_*-{\tilde t})^4},
\ee
which is faster than ${\tilde\rho}_{tot}\sim {1\over ({\tilde
t}_*-{\tilde t})^2}$. Thus the anisotropy problem is still
present, though it is disappeared in the nonminimally-coupled
frame. The explanation might be that the anisotropy $\sigma^2$ is
only a function of $t$, which is not conformally invariant, thus
it's possible that the evolution of the anisotropy in different
frames has different behaviors.
In addition,
as a scenario alternative to inflation, the matter contraction
must be followed by a bounce, i.e. matter bounce
\cite{Brandenberger:2009jq}. However, the implementing of bounce
has been still a challenging issue, e.g.\cite{Battefeld:2014uga}.
Here, we  don't have  to design a mechanism for bounce,
since the universe expands all along, only the reheating is
required. The absence of both the anisotropy problem and the
bounce again suggests that our cosmological background is a
physical expansion.

\section{Discussion}


We are still on the road to detecting the primordial GWs.
We hope that the primordial GWs will bring us the
information about the evolution of the early universe. It is also
possible that the primordial GWs encode physics beyond GR.

We have illustrated a scenario, in which the primordial GWs with
scale-invariant spectrum may emerge from a flat Minkowski space,
by applying the scalar field with nonminimally-derivative coupling
to gravity. In our model, $M_{P,eff}^2\sim (t_*-t)^4 \gg M_P^2$ is
rapidly decreasing during slow expansion, which implies that in
infinite past $M_{P,eff}^2$ is infinite large, so the gravity is
asymptotic-past free.

It is generally thought that the scale-invariance
of primordial GWs spectrum is the significant result of de Sitter
evolution. Thus it is interesting to ask if such a spectrum may
also be produced in other scenarios, which is the reason that we
focus on the scale-invariant spectrum. Moreover, it's also
interesting to construct a scenario with a slightly red GWs
spectrum, which is similar to the observed scalar spectrum, or a
blue-tilted GWs spectrum by letting $M_{P,eff}$ have a different
time dependence. The primordial GWs with slightly blue tilt
$0<n_T<1$, which may appear in some inflation models
e.g.\cite{Piao:2004tq}\cite{Kobayashi:2010cm}\cite{Liu:2013iha}\cite{Cai:2015yza}, might be interesting, since it may boost the
stochastic GWs background at the frequency band of LIGO, as well
as the space-based detectors. Here, we noticed that if
$M_{P,eff}\sim (t_*-t)^{p}$ and $p>1$, we will have $n_T=4-p$, so
$p<4$ means the blue spectrum, and $p>4$ means the red
spectrum. We will come back to this issue in future work.


The slow expansion model proposed in \cite{Piao:2011bz} is
conformally dual to the inflation in Einstein frame, since the
primordial GWs is dominated by its constant mode. Here, our slow
expansion model is actually conformally dual to the matter
contraction, since the primordial GWs is dominated by its growing
mode. However, we have argued that with the matter point of view
our cosmological background is still a physical expansion,
regardless of the frames. In addition, after the slow expansion
ends, we only need a reheating, but not a bounce, since the
universe expands all along. Moveover, maybe more
interestingly, there is no the anisotropy problem.


Though we only focus on a special model of our scenario, the
implementing design is actually universal, i.e., $M_{P,eff}^2\sim
(t_*-t)^4 $ must be satisfied to assure the scale-invariance of
primordial GWs. Thus along the lines in
\cite{Rubakov:2013kaa}\cite{Nishi:2015pta}, we believe that it
could be generally implemented in Horndeski theory and
other theories of modified gravity. It is also possible that such
a Minkowski space is followed by an inflation period,
e.g.\cite{Liu:2014tda}\cite{Kobayashi:2015gga}. In this scenario,
the inflation offers the primordial perturbations responsible for
the large scale structure and CMB fluctuations, while in infinite
past the universe is in flat Minkowski space, which is
geodesic-complete. In addition, it is also interesting to explore
the link of the corresponding scenarios to the string and
supergravity theory.



To conclude, we showed that the primordial GWs with the
scale-invariant spectrum may emerge from a nearly Minkowski space,
in which the gravity is asymptotic-past free. What we would like
to highlight is that exploring the origin of primordial GWs with
different angle of view may offer us a different perspective to
the issues relevant with the early universe scenarios, which in the
meantime might also be significant for an insight into the gravity
physics of primordial universe. Thus the relevant issues are
worthy of studying.

\textbf{Acknowledgments}

This work is supported by NSFC, No. 11222546, 11575188, and the
Strategic Priority Research Program of Chinese Academy of
Sciences, No. XDA04000000.

\appendix

\section{Notebook}\label{appendixa}

Our (\ref{L}) is actually a subclass of so-called Horndeski
theory \cite{Horndeski:1974wa}. Recently, Kobayashi et.al have
calculated the corresponding perturbations
\cite{Kobayashi:2011nu}. Here, we will not involve the details of
calculation and only list the results. In (\ref{action2}) and
(\ref{action3}), we have
\be Q_T=x_1,\ee \be c_T^2={x_4\over x_1}, \ee \be Q_{\cal R}=
3x_1+ {4 x_1^2x_3\over 3x_2^2},\ee \be c_{\cal R}^2 = {2\over
aQ_{\cal R}}\left({a x_1^2\over x_2}\right)^.-{x_4\over Q_{\cal
R}}, \ee and \be x_1 = M_P^2+2f-4Xf_X,\ee  \be x_2 =
2M_P^2H+4f H-16H(X f_X+X^2f_{XX}),\ee \be x_3 =
-9H^2M_P^2-18H^2f+3 X \rho_{eff, X} +
18H^2(7X f_X+16X^2f_{XX}+4X^3f_{XXX}),\ee \be x_4 =
M_P^2+2f, \ee where $\rho_{eff, X}={\cal L}_{1,X}+2X{\cal
L}_{1,XX}$ and $f={\cal M}^8M_P^2/2X^2$. Thus with
Eqs.(\ref{dotphi}) and (\ref{dotH}), $x_i$ can be rewritten as,
respectively, \be x_1=M_p^2+ 5{{\cal M}^8M_P^2 \over X^2}, \ee \be
x_2\simeq \,-30{{\cal M}^8M_P^2 \over X^2}H,\label{x2}\ee \be
x_3\simeq -135 { {\cal M}^8 M_P^2\over X^2}H^2+3X \rho_{eff, X}\ee
\be x_4={{\cal M}^8M_P^2 \over X^2}. \ee


\section{Anisotropy}\label{appendixb}

We begin with the Bianchi-IX metric \cite{Misner:1974qy} \be
ds^2=-dt^2+a^2(t)\sum_{i=1}^{3}e^{2\beta_i(t)}d{x^i}^2, \ee where
$\sum_{i=1}^{3}\beta_i=0$. The evolution of background is given by
\be 3M_p^2H^2=\rho_{tot}+\sigma^2\,, \ee where \be
\sigma^2=\frac{1}{2}\sum_{i=1}^3\dot{\beta}_i^2 \ee is the
anisotropy term. In GR, the equation of motion for $\beta_i$ is
\be \ddot{\beta}_i+3H\dot{\beta}_i=0\,. \label{GRa}\ee Thus
${\dot\beta}_i\sim a^{-3}$ and $\sigma^2\sim a^{-6}$.

In our Lagrangian (\ref{L}), the equation of motion for $\beta_i$
is
\be \lf(1+{5{\cal M}^8\over
{X}^2}\rt)\ddot{\beta}_i+\lf(3H+{15H{\cal M}^8\over
{X}^2}-\frac{10\dot{\phi}\ddot{\phi}{\cal
M}^8}{X^3}\rt)\dot{\beta}_i=0\,, \label{nGRa}
\ee
where $\dot \phi$
is given by Eq.(\ref{dotphi}). When $X\gg {\cal M}^4$, (\ref{L})
corresponds to that in GR, so (\ref{nGRa}) reduces to (\ref{GRa}).
However, in our model, in slowly expanding phase $X\ll {\cal
M}^4$, only considering the dominated part, we have \be {5{\cal
M}^8\over {X}^2}\ddot{\beta}_i-\frac{10\dot{\phi}\ddot{\phi}{\cal
M}^8}{X^3}\dot{\beta}_i=0\,, \ee which gives \be
\frac{\ddot{\beta}_i}{\dot{\beta}_i}=\frac{4}{t_*-t}.\ee  Thus we
have $\dot{\beta}_i\sim(t_*-t)^{-4}$, and the anisotropy is \be
\sigma^2=\frac{1}{2}\sum_i \dot{\beta_i}^2\sim (t_*-t)^{-8}
\label{sigma2}\ee grows with $t$. However, the total energy
density is \be \rho_{tot}\simeq 3M_P^2H^2\sim(t_*-t)^{-18} \ee
grows obviously faster than the anisotropy. Thus in slowly
expanding phase the anisotropy will never dominate the background
if $\sigma^2\ll \rho_{tot}$ initially.

Moreover, it is interesting to check the effect of anisotropy on
the background in Einstein frame. The Bianchi-IX metric is \be
d{\tilde s}^2=-d{\tilde t}^2+{\tilde a}^2\sum_{i=1}^{3}e^{2{\tilde
\beta}_i({\tilde t})}d{x^i}^2, \ee where $\sum_{i=1}^{3}{\tilde
\beta}_i=0$. The equation of motion for $\tilde{\beta}_i$ is
similar to (\ref{GRa}), i.e.,
\be
{d^2\tilde{\beta}_i\over
d\tilde{t}^2}+3\tilde{H}{d\tilde{\beta}_i\over d\tilde{t}}=0\,,
\ee
so we have \be {\tilde\sigma}^2 \sim
1/{\tilde a}^{6}\sim {1\over ({\tilde t}_*-{\tilde t})^4}, \ee
since for the matter contraction ${\tilde a}\sim ({\tilde
t}_*-{\tilde t})^{2/3}$. This result can also be derived from
(\ref{sigma2}) as follows \be {\tilde\sigma}^2= \frac{1}{2}\sum_i
\left({d{\tilde{\beta}_i}\over d{\tilde t}}\right)^2=\frac{1}{2}\sum_i
{\cal A}^2 \dot{\beta_i}^2\sim {1\over (t_*-t)^{12}}\sim {1\over
({\tilde t}_*-{\tilde t})^4}. \ee However, the total energy
density is \be {\tilde\rho}_{tot}\simeq 3M_P^2{\tilde
H}^2\sim({\tilde t}_*-{\tilde t})^{-2} \ee grows slower than the
anisotropy. Thus the anisotropy will eventually dominate the
background. This is the so-called anisotropy problem, which
inevitably appears in the scenario with the matter contraction
phase. However, in certain sense, we think that for our
model, the anisotropy problem appearing in the Einstein frame is
non-physical, since the physical background is actually the
expansion. Similarly, this duality could be used to study certain
features of the chaotic Mixmaster universe.



\end{document}